\documentclass[12pt]{article}

\usepackage{sbc-template}
\usepackage{amsmath}
\usepackage[brazil]{babel}
\usepackage[utf8]{inputenc}
\usepackage{comment}
\usepackage{graphicx,url,breakurl}
\usepackage[breaklinks]{hyperref}

\title{Deep Learning-Based Transfer Learning for Classification of Cassava Disease}

\author{Ademir G. Costa Junior\inst{1}, Fábio S. da Silva\inst{1} \and Ricardo Rios\inst{1}
    \email{\{agdcj.eng20, fssilva, rrios\}@uea.edu.br}
}
\address{Escola Superior de Tecnologia \textendash~Universidade do Estado do Amazonas (UEA)\\ Manaus \textendash~AM \textendash~Brasil
}

\begin{document}

\maketitle

\begin{abstract}
This paper presents a performance comparison among four Convolutional Neural Network architectures (EfficientNet-B3, InceptionV3, ResNet50, and VGG16) for classifying cassava disease images. The images were sourced from an imbalanced dataset from a competition. Appropriate metrics were employed to address class imbalance. The results indicate that EfficientNet-B3 achieved on this task accuracy of 87.7\%, precision of 87.8\%, revocation of 87.8\% and F1-Score of 87.7\%. These findings suggest that EfficientNet-B3 could be a valuable tool to support Digital Agriculture.
\end{abstract}

\begin{resumo}

Esse artigo apresenta um comparativo de desempenho entre quatro arquiteturas de Redes Neurais Convolucionais (EfficientNet-B3, InceptionV3, ResNet50 e VGG16) na classificação de imagens de patologias da mandioca. As imagens foram obtidas de um conjunto de dados desbalanceado de uma competição. Foram utilizadas métricas adequadas para lidar com o desbalanceamento entre classes do conjunto. Os resultados indicam que a EfficientNet-B3 alcançou nessa tarefa acurácia de 87,7\%, precisão de 87,8\%, revocação de 87,8\% e F1-Score de 87,7\%. Isso sugere que o EfficientNet-B3 pode ser uma ferramenta valiosa de apoio para Agricultura Digital.
\end{resumo}

\section{Introdução} \label{sec:Introducao}

De acordo com a Empresa Brasileira de Pesquisa Agropecuária (EMPRABA), o investimento em ciência ao longo das décadas permitiu que o Brasil se posicionasse como uma importante potência mundial na produção de alimentos \cite{embrapa2022}. Para a próxima década, as projeções apontam um significativo potencial de crescimento no setor. Comparado à produção de 2022-2023, espera-se um aumento de 24,1\% até 2032-2033, equivalente a um crescimento anual de 2,4\% \cite{mapa2023}. As projeções para o próximo decênio indicam que o Brasil deve manter ou elevar o padrão que tem sido típico no século atual \textendash~ aumentos da produção agrícola total superiores ao crescimento da área plantada, o que indica um crescimento da produção sem necessidade de expansão da área plantada.

A produção brasileira de mandioca (\textit{Manihot esculenta Crantz}), por exemplo, em 2022 atingiu 18,2 milhões de toneladas em uma área de 1,27 milhões de hectares. Segundo levantamento do Instituto Brasileiro de Geografia e Estatística (IBGE), esse fato consolidou o Brasil como o quinto maior produtor mundial do tubérculo \cite{conab-relatorio2023}. A mandioca assume um papel fundamental na segurança alimentar das regiões tropicais. Sua rusticidade e adaptabilidade a solos ácidos, regimes pluviométricos variados e períodos de seca a tornam ideal para regiões com condições desafiadoras, em que outras culturas podem falhar \cite{farias2006aspectos}.

A mandioca é a base alimentar de mais de 800 milhões de pessoas em países em desenvolvimento. Devido a fatores como doenças e pragas, enfrenta diversos desafios que ameaçam a produção e a segurança alimentar \cite{howeler2013save}. Entre os principais, destacam-se as doenças fúngicas, virais e bacterianas que causam perdas significativas na colheita e na qualidade do tubérculo \cite{calvert2001viruses}.

O diagnóstico preciso e precoce de doenças da mandioca é crucial para o manejo eficaz e a mitigação dos impactos negativos na produção. No entanto, métodos tradicionais de diagnóstico, como a inspeção visual e testes laboratoriais, apresentam diversas limitações como resultados inconsistentes, demora para conclusão do processo e custos elevados \cite{santos2020visao}.

Considerando esse cenário de ameaça à produção de mandioca, a Visão Computacional se apresenta como uma ferramenta promissora na detecção e identificação de pragas e doenças. Especificamente no que diz respeito à agricultura digital, ela se destaca na análise de manifestações perceptuais de doenças, como os sintomas visuais causados por esses patógenos. Através do uso de técnicas avançadas de Inteligência Artificial como Aprendizado Profundo (\textit{Deep Learning} \textendash~DL), especificamente Redes Neurais Convolucionais (\textit{Convolutional Neural Networks} \textendash~CNNs) \cite{lecun2015deep}, é possível a identificação precoce e precisa de patologias na plantação de mandioca e o mapeamento da cobertura do solo \cite{kamilaris2018deep}. Isso viabiliza a adoção de ações mais eficazes e direcionadas \cite{massruha2020agricultura, santos2020visao}.

Nesse contexto, esse artigo apresenta um método de detecção automática de doenças da mandioca, utilizando CNNs para analisar imagens de folhas do tubérculo. São comparadas diferentes arquiteturas de CNNs empregadas na identificação de quatro doenças, a Bacteriose da Mandioca (\textit{Cassava Bacterial Blight} \textendash~CBB), a Doença da Estria Marrom (\textit{Cassava Brown Streak Disease} \textendash~CBSD), a Doença do Mosaico (\textit{Cassava Mosaic Disease} \textendash~CMD) e o Vírus Mosqueado Verde (\textit{Cassava Green Mottle} \textendash~CGM). Além disso, é analisado o desempenho da melhor Arquitetura para a detecção dessas doenças individualmente, pois é possível que uma CNN se comporte melhor na detecção da CBB e não da CBSD.

Esse estudo avalia uma metodologia para a classificação de doenças em mandioca, utilizando técnicas de Aprendizado Profundo e Transferência de Aprendizado. O objetivo principal da pesquisa é desenvolver um modelo robusto e eficiente para o diagnóstico precoce de doenças na cultura da mandioca. Para isso, são comparadas diferentes arquiteturas pré-treinadas de CNNs. O modelo desenvolvido será utilizado como suporte à tomada de decisão no manejo adequado da lavoura no que diz respeito à otimização da produtividade e minimização de perdas.

Esse artigo está organizado da seguinte forma. Na Seção \ref{sec:ArtigosRelacionados}, são apresentados os artigos relacionados à classificação de doenças em plantas utilizando Redes Neurais Artificiais (\textit{Artificial Neural Networks} \textendash~ANNs). Na Seção \ref{sec:ResultExperim}, são apresentadas as configurações experimentais utilizadas, incluindo detalhes sobre o conjunto de dados, modelos, parametrização e métricas de desempenho utilizadas para avaliar os resultados. A Seção \ref{sec:ResultadosDiscussao} aborda os resultados obtidos e a análise comparativa entre os diferentes modelos treinados. Por fim, na Seção \ref{sec:Conclusao}, são apresentadas as considerações finais.

\section{Trabalhos Relacionados} \label{sec:ArtigosRelacionados}

Na literatura, vários trabalhos demonstram a eficácia das ANNs na identificação de patologias em diversas espécies de plantas. Por exemplo, o trabalho \cite{mohanty2016using} investigou o uso de duas arquiteturas de CNNs para classificar 26 doenças em 14 espécies agrícolas, resultando em uma acurácia de 99,35\% no melhor modelo. Os dados utilizados foram obtidos do projeto \textit{Plant Village} \cite{hughes2015open}, que disponibilizou um conjunto de 54.306 imagens.

O trabalho \cite{sangbamrung2020novel} propôs uma abordagem inovadora para a classificação automática de doenças da mandioca. Foram utilizados algoritmos de aprendizado profundo. A metodologia abordou o problema como um caso de classificação binária em duas etapas: \textit{(i)} detecção de doença, diferenciando espécimes saudáveis de espécimes doentes; e \textit{(ii)} classificação específica da doença, classificando a doença específica como \textit{Doença da Raia Marrom da Mandioca} (CBSD) ou outras doenças.

O trabalho \cite{sambasivam2021predictive} introduziu a aplicação de uma CNN para desenvolver um método de aprendizado profundo eficiente e econômico, ou seja, exclui a coleta de amostras e análise laboratorial para detectar infecções nas folhas da mandioca. A abordagem proposta visava resolver um desafio em um conjunto de dados desbalanceado contendo 10.000 imagens classificadas em cinco classes distintas. Para isso, foi utilizada a Técnica de \textit{Oversampling} de Minoria Sintética (\textit{Synthetic Minority Over-sampling Technique} \textendash~SMOTe) para tratar o conjunto de dados desbalanceado e empregaram uma arquitetura CNN composta por três camadas convolucionais e quatro camadas totalmente conectadas no treinamento. Os resultados obtidos demonstraram um F1-Score de até 95\% na classe com o melhor desempenho.

Embora esses trabalho demonstrem um desempenho satisfatório na resolução do problema em foco, é perceptível a ausência de uma análise comparativa abrangente entre diferentes arquiteturas de CNNs para essa tarefa específica. Esse trabalho pretende preencher essa lacuna, apresentando uma comparação detalhada das métricas, utilizando as arquiteturas mais amplamente empregadas em publicações recentes.

\section{Resultados Experimentais} \label{sec:ResultExperim}

\subsection{Configuração dos Experimentos} \label{subsec:ConfExp}

Nesse trabalho, o problema de classificação de doenças da folha da mandioca é considerado como uma tarefa de classificação multiclasse. É utilizanda a técnica de Aprendizado Supervisionado. Isso significa que um conjunto de dados preexistente, com imagens de folhas de mandioca rotuladas com suas respectivas doenças, foi utilizado para treinar os modelos de classificação.

Para o treinamento das arquiteturas de CNNs, foi utilizado um servidor equipado com um processador Intel(R) Core(TM) i7-8700 CPU @ 3,20 GHz, 32 GB de memória principal, 2,4 TB de memória secundária e duas GPUs NVIDIA GeForce GTX 1080 Ti com 11 GB de VRAM cada. O ambiente de programação consistiu na Linguagem de Programação Python 3.12.2 \cite{python3} e no framework PyTorch 2.3.0 \cite{pytorch2019}.

\subsection{Conjunto de dados}\label{sec:ConjDados}

O conjunto de dados utilizado nessa pesquisa compreende 21.367 imagens de folhas de mandioca, cuidadosamente coletadas durante um estudo de campo realizado em Uganda \cite{cassava-leaf-disease-classification}. A maioria das imagens foi capturada por agricultores locais. As imagens foram posteriormente anotadas por especialistas do Instituto Nacional de Pesquisa e Recursos Culturais (NaCRRI), em colaboração com o laboratório de inteligência artificial da Universidade Makerere, em Kampala. As imagens foram cuidadosamente organizadas em cinco classes distintas, cada uma representando uma doença específica da mandioca: Bacteriose da Mandioca (\textit{Cassava Bacterial Blight} \textendash~CBB); Doença da Estria Marrom (\textit{Cassava Brown Streak Disease} \textendash~CBSD); Doença do Mosaico (\textit{Cassava Mosaic Disease} \textendash~CMD); Vírus Mosqueado Verde (\textit{Cassava Green Mottle} \textendash~CGM); e exemplares saudáveis.

\begin{figure}[!htpb]
\centering
\includegraphics[width=\linewidth]{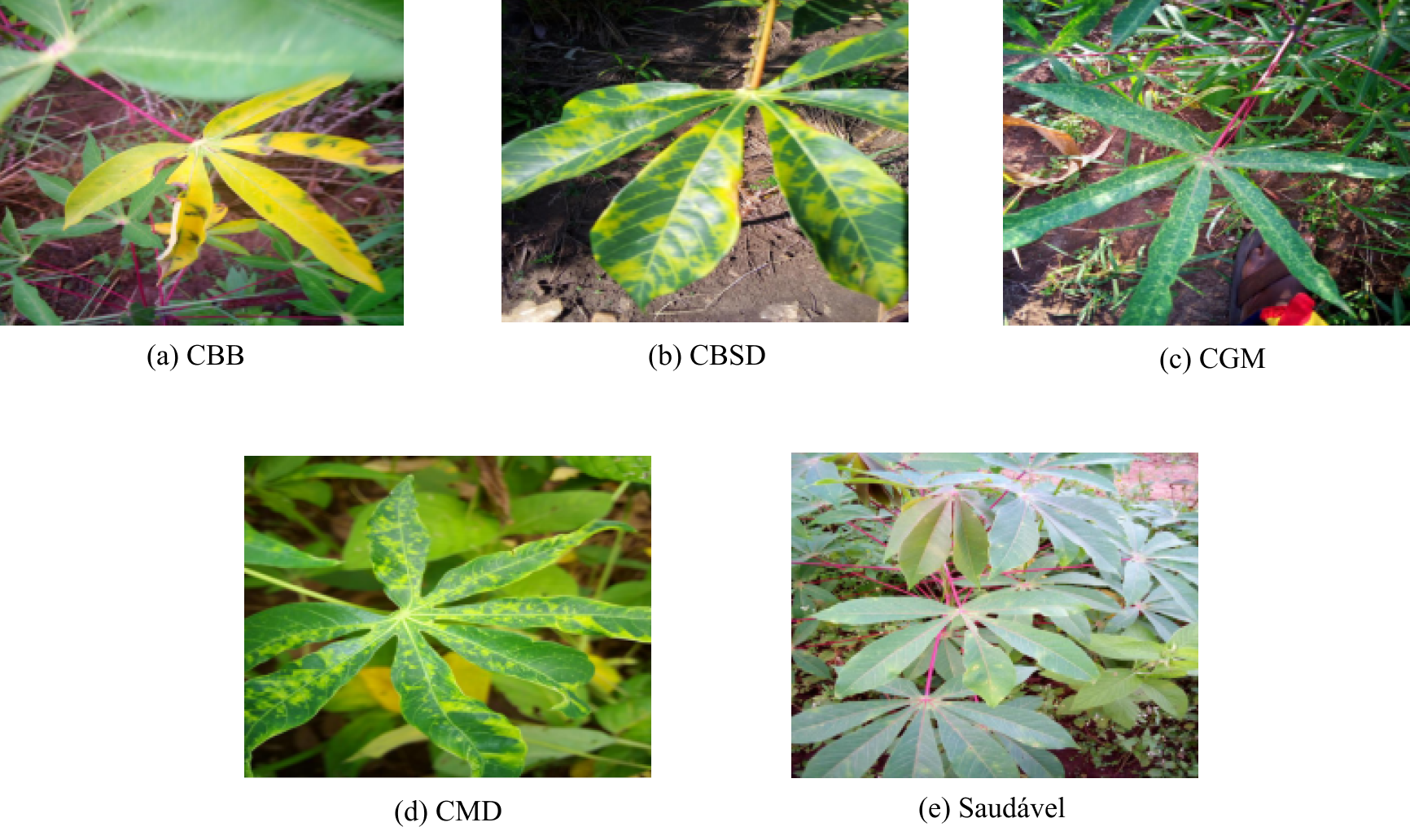}
\caption{Exemplos aleatórios selecionados para cada classe.}
\label{fig:exampleFig1}
\end{figure}

A análise exploratória de dados (EDA) revelou um desbalanceamento significativo entre as classes do conjunto de dados. A classe CMD, por exemplo, apresentou um número de exemplos consideravelmente maior do que a soma de todas as outras classes juntas, sendo responsável por 61.5\% do total de amostras. Classes como a CBB representam apenas 5\%, conforme ilustrado na  Figura \ref{fig:exampleFig1}. Para lidar com esse desafio, nesse trabalho, foram utilizadas métricas especificas para desbalanceamento de classes (Seção \ref{subsec:MetricasAvaliacao}).

\begin{figure}[!htpb]
\centering
\includegraphics[width=.8\linewidth]{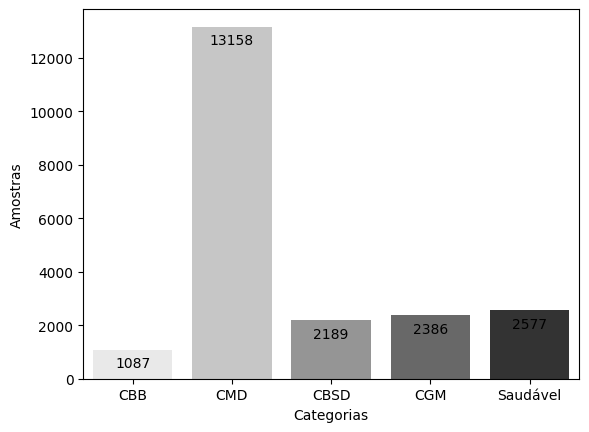}
\caption{Distribuição de amostras na base de dados}
\label{fig:exampleFig1}
\end{figure}

\subsection{Modelos de CNNs} \label{subsec:ModelCNNs}

O método para solução do problema em questão fundamenta-se nos avanços recentes de (\textit{Deep Learning}), nos quais imagens multidimensionais são fornecidas como entrada para CNNs. Essas redes realizam aprendizado automático, massivo e hierárquico dos parâmetros dos filtros extratores de características, capacitando-se, assim, para classificar padrões complexos nos dados de entrada \cite{goodfellow2016deep}.

Foram selecionados os seguintes modelos de CNNs para a solução proposta:

\begin{enumerate}
    \item EfficientNet-B3: proposta por \cite{tan2019efficientnet}, é uma CNN que se destaca pela sua eficiência e desempenho. Essa rede utiliza um método de escalamento composto por coeficientes de largura, profundidade e resolução, para construir arquiteturas computacionalmente eficientes. A EfficientNet-B3 apresenta um equilíbrio entre complexidade e precisão, com 12 camadas principais e é particularmente eficaz em tarefas de visão computacional.

    \item InceptionV3: proposta por \cite{szegedy2016rethinking}, emprega uma arquitetura modularizada com blocos \textit{Inception}, que consistem em convoluções de diferentes tamanhos de filtro, visando aumentar a eficiência computacional e melhorar a precisão. Com 48 camadas convolucionais, é considerada uma das CNNs mais eficientes para tarefas de classificação de imagens.

    \item ResNet50: proposta por \cite{he2016deep}, é uma CNN profunda que utiliza o conceito de camadas residuais para mitigar o problema do desvanecimento do gradiente durante o treinamento. Composta por 50 camadas convolucionais, é amplamente empregada em tarefas de Visão Computacional.

    \item VGG16: proposta por \cite{simonyan2014very}, possui 16 camadas e utiliza filtros convolucionais de tamanho 3x3 para extração de características. Além disso, emprega camadas de \textit{max-pooling} para redução de dimensionalidade, juntamente com camadas finais completamente conectadas para fins de classificação.
\end{enumerate}

\subsection{Pré Processamento das Imagens}\label{subsec:PreProcessamento}

Como mencionado na Seção \ref{sec:ConjDados}, o conjunto de dados apresenta desbalanceamneto entre as classes, com uma classe majoritária predominando significativamente sobre as demais. Para lidar com a situação de classe rara, ou seja, quando uma classe é minoritária, propomos a utilização da estratificação, conforme sugerido por \cite{amaral2016introduccao}. Essa técnica consiste em dividir o conjunto de imagens em subconjuntos, preservando a proporção de classes em cada um. Dessa forma, a divisão do conjunto de dados é apresentada da seguinte maneira:

\begin{itemize}
    \item Treino (70\%): utilizado para treinar o modelo.

    \item Validação (10\%): empregada para avaliar o desempenho do modelo durante o treinamento e evitar o \textit{overfitting}.

    \item Teste (20\%): simula um ambiente real de aplicação, fornecendo uma avaliação imparcial do desempenho final do modelo.
\end{itemize}

Todas as imagens foram redimensionadas para se adequarem às entradas dos diferentes modelos utilizados. Para os modelos ResNet50 e VGG19, as imagens foram ajustadas para 224 x 224 pixels. Para o modelo InceptionV3, as imagens foram redimensionadas para 299 x 299 pixels. Para o modelo EfficientNet-B3, as imagens foram ajustadas para 300 x 300 pixels. Dessa forma, todas as imagens têm as dimensões apropriadas durante o treinamento de cada rede, garantindo a compatibilidade com os respectivos modelos apresentados.

Conjuntamente ao redimensionamento das imagens é aplicado o \textit{Data Augmentation}, uma técnica utilizada em Aprendizado de Máquina e Visão Computacional para aumentar a diversidade dos dados de treinamento. Isso é feito aplicando diversas transformações nos dados originais, criando novas amostras sem a necessidade de coletar dados adicionais \cite{shorten2019survey}.

Para parametrizar o aumento de dados foram realizadas transformações geométricas, de cor e de ruído. Essas transformações incluem normalização dos valores dos pixels, rotação aleatória das imagens, deslocamento horizontal e vertical, transformação de cisalhamento, \textit{zoom} aleatório, inversão horizontal e ajuste de brilho. É importante destacar que essas transformações são aplicadas apenas ao conjunto de treinamento. Nos conjuntos de teste e validação, apenas a normalização dos valores dos pixels é aplicada.

\subsection{Transferência de Aprendizado} \label{subsec:TransAprendizado}

Treinar uma CNN para uma tarefa específica pode ser um processo computacionalmente caro. Essa tarefa exige grandes conjuntos de dados e poder de processamento significativo. A Transferência de Aprendizado \cite{pan2009survey} oferece uma solução para contornar essa dificuldade. Essa técnica utiliza um modelo pré-treinado em um problema similar como ponto de partida para o treinamento da rede. Dessa forma, é possível otimizar o treinamento de redes convolucionais, especialmente quando recursos computacionais são limitados.

Para iniciação dos pesos dos modelos foi utilizado o pré-treinamento do ImageNet \cite{russakovsky2015imagenet}, aproveitando os pesos das camadas genéricas para iniciar o treinamento para acelerar o processo de treinamento.

\subsection{Construção do Modelo} \label{subsec:ConstModelo}

Para todas as arquiteturas de CNNs selecionadas (Secão \ref{subsec:ModelCNNs}), os pesos foram iniciados conforme descrito na Seção \ref{subsec:TransAprendizado}. O treinamento foi conduzido com um limite máximo de 50 épocas. Adotou-se a técnica de \textit{Early Stopping}, com uma paciência de 5 épocas, interrompendo o treinamento caso a perda no conjunto de validação não melhore em 5 épocas consecutivas. Além disso, foi empregada a técnica de \textit{Model Checkpoint} para monitorar a perda no conjunto de validação e persistir os pesos que proporcionassem melhor generalização.

O tamanho do lote (\textit{batch size}) foi estabelecido em 32, o que significa que o algoritmo treinou os modelos em blocos de 32 amostras, até utilizar todo o conjunto de dados. Quando todos os blocos são passados pelo modelo, encerra-se uma época.

A taxa de aprendizado inicial foi definida em $10^{-3}$ com o uso do otimizador Adam \cite{kingma2014adam}, conhecido por sua capacidade de lidar com problemas de convergência em treinamentos de redes neurais profundas e por sua robustez em relação a variações na taxa de aprendizado.

A técnica de identificação de platô (\textit{ReduceLROnPlateau}) foi implementada com o objetivo de otimizar a taxa de aprendizado do modelo. Essa estratégia assegura que, caso o modelo não apresente melhorias em suas métricas (a \textit{loss} do conjunto de validação) por duas épocas consecutivas, a taxa de aprendizado seja automaticamente reduzida em um fator de 0,1. Isso evita a estagnação e promove um treinamento mais eficiente.

Para o cálculo da \textit{loss}, optou-se pelo uso do critério Perda de Entropia Cruzada (\textit{Cross-Entropy Loss}), também conhecido como Perda Logarítimica, disponível no Pytorch. Essa função de perda é usada para encontrar a solução ideal ajustando os pesos de um modelo de aprendizado de máquina durante o treinamento. O objetivo é minimizar o erro entre os resultados reais e previstos. Assim, uma medida mais próxima de 0 (zero) é sinal de um bom modelo, enquanto uma medida mais próxima de 1 (um) é sinal de um modelo de baixo desempenho.

A configuração das redes pré-treinadas é fundamental, especialmente considerando que foram originalmente treinadas no conjunto de dados \textit{ImageNet}, que consiste em 1000 classes de saída possíveis. Para adaptar o modelo à tarefa de classificação de doenças da mandioca, que possui cinco classes de saída, é necessário realizar modificações na arquitetura. Inicialmente, as camadas finais do modelo são removidas para permitir a inclusão de camadas adequadas a essa classificação. Adiciona-se uma camada densa com cinco neurônios e ativação \textit{softmax}, que converte as saídas da camada densa em probabilidades. Cada neurônio na camada de saída representa a probabilidade da imagem pertencer a uma das classes especificadas e considera-se que a classe com maior probabilidade é a predição do modelo.

\subsection{Métricas de Avaliação} \label{subsec:MetricasAvaliacao}

A avaliação da eficácia dos modelos selecionados requer métricas ponderadas adequadas ao desbalanceamento do conjunto de dados. Para isso, foram selecionadas métricas como acurácia, precisão, revocação e F1-Score, juntamente com a análise da matriz de confusão.

\begin{equation} \label{eq:01}
\text{Acurácia} = \frac{1}{|N|} \sum_{n\in N} \left( \frac{\text{TP}_n + \text{TN}_n}{\text{TP}_n + \text{TN}_n + \text{FP}_n + \text{FN}_n}\right)
\end{equation}

\begin{equation} \label{eq:02}
\text{Precisão} = \frac{1}{|N|} \sum_{n\in N} \left( \frac{\text{TP}_n}{\text{TP}_n + \text{FP}_n}\right)
\end{equation}

\begin{equation} \label{eq:03}
\text{Revocação} = \frac{1}{|N|} \sum_{n\in N} \left( \frac{\text{TP}_n}{\text{TP}_n + \text{FN}_n} \right)
\end{equation}

\begin{equation} \label{eq:04}
\text{F1-Score} = \frac{2 \cdot \text{Precisão} \cdot \text{Revocação}}{(\text{Precisão} + \text{Revocação})}
\end{equation}

Nas equações (\ref{eq:01})-(\ref{eq:03}), em que \textit{N} denota o conjunto de classes do problema, as abreviações representam quatro possíveis resultados em uma análise de classificação binária. Explicitamente, \textit{True Positive} (TP) quantifica as instâncias corretamente identificadas como pertencentes à classe positiva; \textit{True Negative} (TN) reflete as predições corretas para a classe negativa; \textit{False Positive} (FP) representa os casos erroneamente classificados como positivos quando deveriam ser negativos; e \textit{False Negative} (FN) representa os casos em que as predições indicam negatividade equivocadamente, quando deveriam ser positivas.

\section{Resultados e Discussão} \label{sec:ResultadosDiscussao}

Os experimentos foram conduzidos de acordo com a metodologia proposta na Seção \ref{sec:ResultExperim} e os resultados foram registrados na Tabela \ref{tab:cnn_metrics}. Nessa seção, são apresentados os resultados e discute-se o uso das estratégias de treinamento empregadas, bem como avalia-se o desempenho da classificação de cada CNN escolhida, conforme as métricas descritas na Seção \ref{subsec:MetricasAvaliacao}.

Para o comparativo principal entre os modelos, utilizou-se a métrica \textit{F1-Score}, pois, além de equilibrar Precisão e Revocação, essa métrica é especialmente útil em cenários com conjuntos de dados desbalanceados, oferecendo uma avaliação mais precisa em situações em que algumas classes são subrepresentadas.

\begin{table}[!htpb]
\centering
\caption{Resultados obtidos para as Arquiteturas Selecionadas}
\label{tab:cnn_metrics}
\begin{tabular}{lcccccc}
\hline
\textbf{Arquitetura} & \textbf{Parâmetros (M)} & \textbf{Acurácia} & \textbf{Precisão} & \textbf{Revocação} & \textbf{F1-Score} \\
\hline
EfficientNet-B3 & 12 & 0.877 & 0.878 & 0.877 & 0.877 \\
InceptionV3 & 23.8 & 0.866 & 0.867 & 0.866 & 0.866 \\
ResNet50 & 25.6 & 0.863 & 0.862 & 0.863 & 0.862 \\
VGG16 & 138 & 0.615 & 0.378 & 0.615 & 0.468 \\
\hline
\end{tabular}
\end{table}

Com exceção da VGG16, todas as arquiteturas alcançaram resultados acima de 86\% em todas as métricas calculadas sobre o conjunto de testes. Dentre elas, destaca-se positivamente a EfficientNet-B3, que obteve um \textit{F1-Score} de 87,7\%. A VGG16 apresentou um desempenho inferior devido à sua incapacidade de capturar padrões e características relevantes em um conjunto de dados tão desbalanceado. Como resultado, essa arquitetura apresentou um comportamento em que a predição era sempre a classe majoritária para todos os exemplos. Por isso sua acurácia é igual à proporção de casos da classe majoritária (CMD).

Entre os diversos fatores que justificam a ineficiência da VGG16 em comparação com outros modelos, destaca-se sua quantidade excessiva de parâmetros. A VGG16 possui até 10 vezes mais parâmetros do que as melhores arquiteturas, o que aumenta a propensão ao \textit{overfitting}. Por outro lado, modelos como EfficientNet-B3, InceptionV3 e ResNet50 não apenas possuem menos parâmetros, mas também são projetados para serem mais eficientes computacionalmente, permitindo alcançar melhores resultados com menos recursos. O EfficientNet-B3 incorpora técnicas avançadas de regularização, como o escalonamento composto, que envolve o aumento simultâneo de largura, profundidade e resolução da rede. Isso permite que a rede receba imagens de maior resolução, com mais camadas convolucionais e um número maior de canais em cada camada, capturando características mais complexas dos dados.

\begin{figure}[!htpb]
\centering
\includegraphics[width=.7\linewidth]{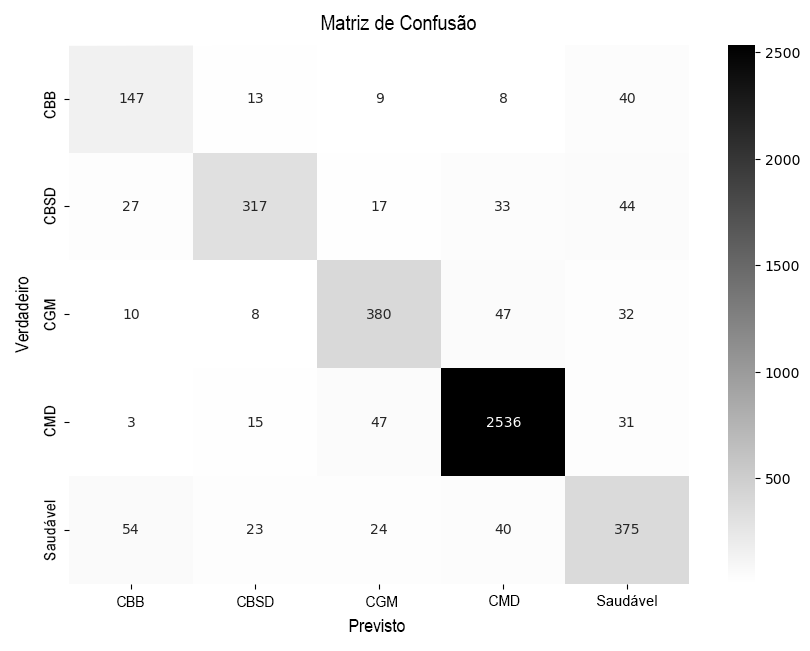}
\caption{Matriz de Confusão da EfficientNet-B3}
\label{fig:Fig2}
\end{figure}

Devido aos resultados superiores da arquitetura EfficientNet-B3, decidiu-se explorar mais detalhadamente seus resultados. Observando a Tabela \ref{tab:classif_report} e a Matriz de Confusão apresentada na Figura \ref{fig:Fig2}, nota-se que a arquitetura foi capaz de classificar todas as classes com mais de 70\% de precisão, exceto a classe CBB. Essa classe tem a menor quantidade de amostras do conjunto de dados, representando cerca de 5\% do total de amostras. A classe CBB teve o pior desempenho. Destaca-se a classificação da classe CMD, a classe majoritária do problema, com todas as métricas acima de 95\%. Apesar de as classes CBSD, CGM e os exemplares saudáveis também serem classes raras no conjunto, os resultados para elas foram superiores, chegando a atingir até 84.3\% de precisão. Isso demostra que é possível atingir bons resultados apesar do conjunto desbalanceado.

\begin{table}[!h]
\centering
\caption{Relatório de Classificação}
\label{tab:classif_report}
\begin{tabular}{lcccc}
\hline
\textbf{Classe} & \textbf{Precisão} & \textbf{Revocação} & \textbf{F1-score} & \textbf{Suporte} \\
\hline
CBB & 0.610 & 0.677 & 0.642 & 217 \\
CBSD & 0.843 & 0.724 & 0.779 & 438 \\
CGM & 0.797 & 0.797 & 0.797 & 477 \\
CMD & 0.952 & 0.964 & 0.958 & 2632 \\
Saudável & 0.718 & 0.727 & 0.723 & 516 \\
\hline
\end{tabular}
\end{table}

\section{Conclusões} \label{sec:Conclusao}

Esse trabalho teve como objetivo realizar um estudo comparativo do desempenho de diferentes arquiteturas de Redes Neurais Convolucionais utilizando a técnica de \textit{Transfer Learning}. O objetivo foi encontrar um modelo eficaz na classificação de patologias da Mandioca, otimizando recursos computacionais e eliminando a necessidade de desenvolvimento de um modelo específico para essa tarefa.

Durante o desenvolvimento desse estudo, foi observado que as Redes Neurais Convolucionais podem ser ferramentas úteis na classificação dessas patologias. Destaca-se o EfficientNet-B3, que alcançou um \textit{F1-Score} de 87,7\% e até 95\% de acurácia na classe com maior número de acertos, demonstrando sua capacidade de identificar corretamente a maioria das patologias apresentadas.

Para futuros trabalhos, planeja-se explorar o impacto de diferentes hiperparâmetros, como, por exemplo, o otimizador RMSProp \cite{tieleman2017divide} ou variantes do Adam. Planeja-se também explorar técnicas avançadas de otimização de hiperparâmetros, como, por exemplo, a busca em larga escala. Além disso, pretende-se avaliar o efeito do balanceamento do conjunto de dados através de técnicas, como replicação de imagens, ou geração de imagens sintéticas para classes minoritárias usando métodos, como SMOTE \cite{chawla2002smote} ou ADASYN \cite{he2008adasyn}.

\section{Agradecimentos}

Os autores agradecem pelo apoio fornecido pelo Laboratório de Sistemas Inteligentes (LSI) da Universidade do Estado do Amazonas (UEA), na pessoa da Professora do Núcleo de Computação (NuComp) da Escola Superior de Tecnologia (EST). Dra. Elloá B. Guedes.

\end{document}